\documentclass[11pt,a4paper]{article}

\usepackage{amsfonts,amssymb,amsthm}
\usepackage{cite} 
\usepackage[T2A]{fontenc}
\usepackage[cp1251]{inputenc}
\usepackage[english]{babel}
\usepackage{amsmath}%
\usepackage[unicode]{hyperref}
\usepackage[left=1.5cm,right=1.5cm,
    top=2cm,bottom=3cm,bindingoffset=0cm]{geometry}

\allowdisplaybreaks
\numberwithin{equation}{section}

\title{\bf On integrable reductions of two-dimensional Toda-type lattices.}
\author{\bf I.T.Habibullin, A.U.Sakieva
}

\date{} 

\begin{document}
\maketitle



\abstract {The article considers lattices of the two-dimensional Toda type, which can be interpreted as dressing chains for spatially two-dimensional generalizations of equations of the class of nonlinear Schr\"odinger equations. The well-known example of this kind of  generalization is the Davey-Stewartson equation. It turns out that the finite-field reductions of these lattices, obtained by imposing cutoff boundary conditions of an appropriate type, are Darboux integrable, i.e., they have complete sets of characteristic integrals. An algorithm for constructing complete sets of characteristic integrals of finite field systems using Lax pairs and Miura-type transformations is discussed.} 

\vspace{0.5cm}

\textbf{Keywords:} {integrability in sense of Darboux, generating function of integrals, Lax pairs, Liouville type equations, Volterra chain}

\vspace{0.5cm}

\large

\section{Introduction} 

Nonlinear equations of the two-dimensional Toda lattice type are closely related to spatially two-dimensional generalizations of the nonlinear Schr\"odinger equation \cite{MikhailovYamilov}. First of all, this is the Benney-Roskes (Davey-Stewartson) equation, obtained in the works \cite{BenneyRoskes}, \cite{DaveyStewartson}. A detailed discussion of the connection between nonlinear lattices and integrable systems of nonlinear partial differential equations can be found in the works \cite{MikhailovYamilov}, \cite{ShabatYamilov}, \cite{Ueno}.

As it is known, the sequence of Laplace invariants $\left\{h_j\right\}$ of a linear partial differential equation of hyperbolic type is described by a nonlinear equation of the form \cite{Laplace}
\begin{equation}\label{Darboux}
(\log h_j)_{xy}+2h_j-h_{j+1}-h_{j-1}=0
\end{equation}
which, according to modern terminology, is one of the versions of the Toda lattice. It is interesting that under imposing termination conditions of the form $h_0=0$, $h_{N+1}=0$, the lattice \eqref{Darboux} is reduced to the following system of hyperbolic type equations
\begin{eqnarray}
&&(\log h_{1})_{xy}+2h_{1}-h_{2}=0,\nonumber\\
&&(\log h_{2})_{xy}+2h_{2}-h_{1}-h_{3}=0, \nonumber \\
&&............................................\nonumber\\
&&(\log h_{N})_{xy}+2h_{N}-h_{N-1}=0 \nonumber
\end{eqnarray} 
for which in the 19th century Darboux found an explicit formula for general solution (see, for instance, \cite{Ganzha}). Another example where a finite-field reduction of an arbitrary size of a nonlinear three-dimensional lattice is solved explicitly is considered in \cite {LSS81}. There the two-dimensional Volterra coupled system is investigated
\begin{equation}\label{2Dvolterra}
a_{n,y}=a_n(b_n-b_{n+1}), \quad b_{n,x}=b_n(a_{n-1}-a_{n}),
\end{equation}
introduced in relation to environmental problems in \cite{Volterra31} (in particular, to study the dynamics of species coexistence). In physical applications, this equation arises when studying the fine structure of the spectra of Langmuir waves in plasma and describes, under suitable boundary conditions, the propagation of the spectral packet of Langmuir oscillations against the background of thermal noise (see \cite{Zakharov73}, \cite{Zakharov74}).

Thus, we observe an intriguing fact: some of the Toda-type lattices in 3D admit a hierarchy of reductions that have increased integrability such as exact solvability. Initiated with this observation we suggested a classification algorithm of the integrable 3D lattices based on the existence of a hierarchy of the reductions satisfying the more weak requirement than exact solvability. We assume that the reduced systems have complete sets of the characteristic integrals corresponding to each characteristic direction (see \cite{Habibullin13},  \cite{HabibullinPoptsova18}). Recall that systems of the hyperbolic type equations admitting complete sets of integrals for each direction are called integrable in the sense of Darboux (for the details see, for instance, monograph \cite{ZMHS}). An effective criterion for Darboux integrability of a PDE system of hyperbolic type, based on the concept of characteristic Lie algebras, was proposed by A.B. Shabat in 1980 in his report at the conference dedicated to the memory of I.G. Petrovsky. An important statement is proved that a system of the hyperbolic type equations is integrable in the sense of Darboux if and only if its characteristic Lie algebras corresponding to both characteristic directions are of finite dimension (see \cite{ShabatYamilov81}, \cite{ZMHS}).

In the article \cite{Habibullin13} a problem of the complete description of the integrable equations of the form
\begin{equation}  \label{todatype}
u_{n,xy} = f(u_{n+1},u_{n},u_{n-1}, u_{n,x},u_{n,y})
\end{equation}
by using the method of the characteristic Lie algebras is considered. Later in \cite{HabibullinPoptsova18}
the problem is formulated and discussed to describe all equations of the form \eqref{todatype} for which there exists a pair of functions $f_1=f_1\left(u_1,u_2,u_{1,x},u_{1,y}\right)$ and $f_2=f_2\left(u_m,u_{m-1},u_{m,x},u_{m,y}\right)$ such that for any integer $m\geq2$ the reduced system of the hyperbolic type equations
\begin{equation}\label{reduc}
\begin{aligned}
&u_{1,xy}=f_1\left(u_1,u_2,u_{1,x},u_{1,y}\right),\\
&u_{j,xy}=f\left(u_{j+1},u_j,u_{j-1},u_{j,x},u_{j,y}\right), \quad 1<j<m, \\
&u_{m,xy}=f_2\left(u_m,u_{m-1},u_{m,x},u_{m,y}\right)
\end{aligned}
\end{equation}
is integrable in the sense of Darboux.

For lattices having the following particular quasilinear form
\begin{equation}\label{0}
u_{n,xy}=A_1u_{n,x}u_{n,y}+A_2u_{n,x}+A_3u_{n,y}+A_4    
\end{equation}
where the coefficients are smooth functions of the dynamical variables $A_i=A_i(u_{n+1},u_n,u_{n-1})$ for $i=1,2,3,4$, 
the classification problem mentioned above was completely solved in the works \cite{HabibullinPoptsova18}, \cite{HabibullinKuznetsova20}, \cite{Kuznetsova19}. This requirement was motivated by the fact that all examples of integrable chains in 3D known in the literature depend linearly on each of the first derivatives. Here we give the complete list of the lattices in the class \eqref{0} which pass the test formulated above:
\begin{itemize}
\item[$(E1)$] $u_{n,xy} = e^{u_{n+1} - 2 u_n + u_{n-1} },$
\item[$(E2)$] $u_{n,xy} = e^{u_{n+1}} - 2 e^{u_n} + e^{u_{n-1}},$
\item[$(E3)$] $u_{n,xy} = e^{u_{n+1}-{u_n}} -  e^{u_n-u_{n-1}},$
\item[$(E4)$] $u_{n,xy} = \left(u_{n+1} - 2 u_n + u_{n-1}  \right) u_{n,x}, $
\item[$(E5)$] $u_{n,xy} = \left(e^{u_{n+1}-{u_n}} -  e^{u_n-u_{n-1}}\right)u_{n,x},$
\item[$(E6)$] $u_{n,xy}=\alpha_nu_{n,x}u_{n,y}, \quad \alpha_n = \frac{1}{u_n - u_{n-1}} - \frac{1}{u_{n+1}-u_n}$
\item[$(E7)$] $u_{n,xy} = \alpha_n(u_{n,x} + u^2_n - 1)(u_{n,y} + u^2_n - 1) - 2 u_n(u_{n,x}+u_{n,y}+u^2_n - 1).$
\end{itemize}

The classification problem in the general case of \eqref{todatype} remains open. Alternative approaches to the problem of classifying integrable lattices in 3D are discussed, for example, in the works \cite{Ferapontov15}, \cite{Ferapontov20}.

Equations $(E1)$-$(E6)$ are well-known integrable models (see, for instance, \cite{ShabatYamilov}). Equation $(E7)$ was obtained during the classification process in \cite{HabibullinPoptsova18}. The Lax pair for it was found in \cite{Kuznetsova21}.
For the lattices $(E6)$, $(E7)$ the reductions of the form \eqref{reduc} are investigated in \cite{HabibullinPoptsova17} and correspondingly in \cite{HabibullinPoptsova18} admitting finite-dimensional characteristic Lie algebras for arbitrary $m$. However in these articles the integrals are not presented.
Reductions of the form \eqref{reduc} for the lattice $(E1)$ coincide with the exponential type systems related to the Cartan matrices of the simple Lie algebras of the series A. As it was shown in \cite{ShabatYamilov81} they are certainly Darboux integrable for arbitrary $m$.
In the case of the lattice $(E3)$ the Darboux integrable reductions are studied in \cite{Smirnov15}, where a generating function for the characteristic integrals for the reduced systems are obtained. 

Characteristic Lie algebras provide an effective (see \cite{ZMHS},  \cite{GZH12}), but rather labor-intensive method for constructing integrals, so it is not  convenient for large values of $m$. Various alternative approaches to the problem of constructing characteristic integrals, including Laplace's cascade method in scalar and matrix versions, were previously used in the works \cite{Smirnov15}, \cite{Demskoi10}, \cite{Anderson}, \cite{Zhiber}, \cite{Zhiber04}, \cite{HabibullinKhakimova21}. 
The purpose of the present article is to develop a method for constructing characteristic integrals based on the use of the Lax pair and the Miura type transformations and then apply it to some key  equations of the list above.

It is well known that in the case of the soliton systems, in order to find the densities of the integrals of motion, one can use the formal asymptotic expansion of the Lax operator eigenfunction  in powers of the spectral parameter. This technique is not suitable for searching for characteristic integrals, since Lax pairs for Darboux integrable systems do not contain any spectral parameter.

Let us briefly discuss on the content of the article. In the second section, the definition of the complete set of independent characteristic integrals of a system of hyperbolic type differential equations is given, and a convenient criterion for the completeness is formulated. Using the Liouville equation as an example, it is shown that the presence of a complete set of integrals ensures separation of the variables, which, in turn, reduces the problem of integrating PDEs to the problem of finding explicit solutions to ODEs (see Remark below).
In Section 3, we briefly discuss the work of S. V. Smirnov \cite{Smirnov15}, where generating functions of characteristic integrals for several finite-field versions of the Toda lattice $(E3)$ are presented. The approach of \cite{Smirnov15} is closely related to the sequence of the Laplace transformations of a linear equation of hyperbolic type with variable coefficients. We modified this approach and adapted it to the case of equations that are not directly related to the Laplace cascade method. In sections 4-6 we apply this algorithm for finding integrals to specific systems that are reductions of the lattices $(E6)$, $(E7)$ and, correspondingly, the coupled system \eqref{2Dvolterra}. In all three cases generating functions are given, allowing one to construct complete sets of characteristic integrals for reductions of the form \eqref{reduc} of arbitrary order $m$.

\section{Complete sets of characteristic integrals}

Let us consider a system of the hyperbolic type equations of the form
\begin{equation}\label{system}
u_{i,xy}=F_i(u_1,\dots u_N, u_{1,x},\dots u_{N,x},u_{1,y},\dots u_{N,y})\quad i=1,2,\dots N,
\end{equation}
where $F_i$ are analytic functions defined in some region of the space $\textbf{С}^{3N}$.

We recall some basic definitions (see, for example, \cite{ZMHS}). Smooth function of the form
$$I=I(u_1,\dots u_N, u_{1,x},\dots u_{N,x},u_{1,xx},\dots u_{N,xx},\dots, u_{1,r},\dots u_{N,r}),$$
where the notation is accepted $u_{j,s}:=\frac{\partial^s}{\partial x^s}u_j$, is called a $y$-integral of the order $r$ of the system \eqref{system}, if the condition is satisfied $D_yI=0$, where the derivative with respect to the variable $y$ is evaluated by means of the system \eqref{system}. In other words to look for the $y$-integral we have a relation
$$D_yI=\sum_i(u_{i,y}\frac{\partial}{\partial u_i}+ F_i \frac{\partial}{\partial u_{i,x}}+ D_x(F_i)\frac{\partial}{\partial u_{i,xx}}+\cdots)I=YI=0.$$
  
A set of $y$-integrals $\left\{I_i\right\}^{i=N}_{i=1}$ having the orders $r_i$ for the system of equations \eqref{system} constitutes a complete set of independent $y$-integrals, if the condition holds
\begin{equation}\label{detdifInt}
\det\left(\frac{\partial I_j}{\partial u_{i,m}}\right)=
\begin{vmatrix}
 \frac{\partial I_1}{\partial u_{1,r_1}} & \frac{\partial I_1}{\partial u_{2,r_1}} & \ldots & \frac{\partial I_1}{\partial u_{N,r_1}}\\
\ldots&\ldots&\ldots&\ldots\\
  \frac{\partial I_N}{\partial u_{1,r_N}} & \frac{\partial I_N}{\partial u_{2,r_N}} & \ldots & \frac{\partial I_N}{\partial u_{N,r_N}}
\end{vmatrix}\neq0.
\end{equation}

Let's formulate an effective algebraic criterion (see \cite{ZMHS}) for the existence of a complete set of integrals of the system \eqref{system}.

{\bf Theorem 1}. {\it System of equations \eqref{system}
admits a complete set of independent $y$-integrals if and only if the characteristic Lie algebra over the ring of locally analytic functions of the variables $\bar u_y, \bar u, \bar u_x, \bar u_{xx},\dots$, where $\bar u=(u_1,u_2,\dots, u_N)$, generated by the characteristic operators
$$X_j=\frac{\partial}{\partial u_{j,y}},\, j=1,\dots N;\qquad Y=\sum_iu_{i,y}\frac{\partial}{\partial u_i}+ F_i \frac{\partial}{\partial u_{i,x}}+ D_x(F_i)\frac{\partial}{\partial u_{i,xx}}+\cdots$$
is of finite dimension.} 

Integrals and characteristic algebra in the direction $x$ are defined similarly.
As an illustrative example, we consider the well-known Liouville equation
\begin{equation}\label{L}
u_{xy}=e^u.
\end{equation}
Differentiating it with respect to $x$ and using the equation \eqref{L} once again, we find $u_{xxy}=u_ x u_{xy}$. Next, we integrate it over $y$ and arrive at the relation
\begin{equation}\label{yint}
u_{xx}-\frac{1}{2}u^2_x=h(x),
\end{equation}
where $h(x)$ is an arbitrary function. Similarly we have
\begin{equation}\label{xint}
u_{yy}-\frac{1}{2}u^2_y=g(y).
\end{equation}
Now it's clear that the functions $I=u_{xx}-\frac{1}{2}u^2_x$ and $J=u_{yy}-\frac{1}{2}u^2_y$ are, respectively, $y-$ and $x-$integrals. Due to the presence of the complete sets of integrals, the variables in the Liouville equation are separated in such a way that the problem of integrating the equation is reduced to finding a joint solution of two ordinary differential equations. From the relations \eqref{yint}, \eqref{xint} it is easy to derive the formula for the general solution of the equation \eqref{L}, found by Liouville:
\begin{equation*}
u\left(x,y\right)=\ln\frac{2p'(x)q'(y)}{\left(p(x)+q(y)\right)^2},
\end{equation*}
where $p=p(x)$, $q=q(y)$ are arbitrary functions, and the prime over the letter denotes the derivative.

{\bf Remark.} This example clearly indicates that the presence of Darboux-integrable reductions can be successfully used to construct explicit particular solutions of the original three-dimensional Toda-type chain by first finding a solution to the reduced system and then extending it to all values of the discrete variable (see \cite{Kuznetsova23}, \cite{KHKh23}).

\section{On the algorithm for constructing characteristic integrals by using the Lax pair.}

In mathematical physics, the two-dimensional Toda lattice equation \eqref{Darboux} is often used in the following form
\begin{equation}\label{Toda}
u_{n,xy} = e^{u_{n+1}-{u_n}} -  e^{u_n-u_{n-1}},
\end{equation}
where the dependent variables are related as follows $h_j=\exp{(u_j-u_{j-1})}$. The lattice \eqref{Toda} is an integrable equation in the sense that it admits the Lax representation, i.e. expresses the necessary and sufficient condition for the compatibility of an overdetermined system of linear equations of the form \cite{Mikhailov79}
\begin{equation}\label{LaxToda}
\psi_{n,x}= u_{n,x}\psi_{n}+\psi_{n+1}, \quad \psi_{n,y}=-  e^{u_n-u_{n-1}}\psi_{n-1}.
\end{equation}

The Toda lattice \eqref{Toda} admits several types of finite field reductions that are Darboux integrable systems. Here we will focus on the reductions corresponding to the Cartan matrices of simple Lie algebras of the series A: 
\begin{eqnarray}
&&u_{1,xy}=e^{u_{2}-u_{1}},\nonumber\\
&&u_{2,xy}=e^{u_{3}-u_2}-e^{u_{2}-u_{1}}, \nonumber \\
&&......................................... \label{rToda}\\
&&u_{N-1,xy}=e^{u_{N}-u_{N-1}}-e^{u_{N-1}-u_{N-2}}, \nonumber \\
&&u_{N,xy}=-e^{u_{N}-u_{N-1}}. \nonumber
\end{eqnarray} 
These reductions are obtained from \eqref{Toda} by imposing the termination conditions $u_0=\infty$ and $u_{N+1}=-\infty$.

It is well known that the system \eqref{rToda} admits complete sets of characteristic integrals in both characteristic directions $x$ and $y$ (see  \cite{Smirnov15}). In the mentioned work S.V. Smirnov found the generating functions of characteristic integrals of generalized Toda lattices corresponding to simple Lie algebras of the series $A$, $B$ and $C$, using the properties of a sequence of the Laplace transformations of a linear partial differential equation of hyperbolic type and thereby proved the Darboux integrability of these systems. Below, in this section, we slightly modify the scheme suggested in  \cite{Smirnov15}, focusing on the Lax pair in order to then apply it to a more general situation that does not have a direct connection with iterations of the Laplace transformation.

The Lax pair for the system \eqref{rToda} is easily derived from the system \eqref{LaxToda}. The first Lax equation has the form of a system of ODEs in $x$
\begin{eqnarray}
&& \psi_{1,x}=u_{1,x}\psi_1+\psi_2,\nonumber\\
&& \psi_{2,x}=u_{2,x}\psi_2+\psi_3, \nonumber \\
&&......................................... \label{LaxxrToda}\\
&&\psi_{N-1,x}=u_{N-1,x}\psi_{N-1}+\psi_{N}, \nonumber \\
&&\psi_{N,x}=u_{N,x}\psi_{N}. \nonumber
\end{eqnarray} 
The second Lax equation is defined similarly
\begin{eqnarray}
&& \psi_{1,y}=0,\nonumber\\
&& \psi_{2,y}=-e^{u_2-u_1}\psi_{1}, \nonumber \\
&&......................................... \label{LaxyrToda}\\
&&\psi_{N-1,y}=-e^{u_{N-1}-u_{N-2}}\psi_{N-2}, \nonumber \\
&&\psi_{N,y}=-e^{u_{N}-u_{N-1}}\psi_{N-1}. \nonumber
\end{eqnarray} 

The system of equations \eqref{LaxxrToda} obviously implies the relation
\begin{equation}
\label{relationN}
\psi_N=(D_x-u_{N-1,x})(D_x-u_{N-2,x})\dots(D_x-u_{2,x})(D_x-u_{1,x})\psi_1.
\end{equation}
Hence due to the fact that $(D_x-u_{N,x})\psi_N=0$ the function $\psi_1$ is a joint solution of two ordinary differential equations
\begin{equation}
\label{relation2}
B\psi_1=0, \quad\frac{\partial}{\partial y}\psi_1=0,
\end{equation}
where
\begin{equation}
\label{operatorBproduct}
B=(D_x-u_{N,x})(D_x-u_{N-1,x})\dots(D_x-u_{2,x})(D_x-u_{1,x}).
\end{equation}
Let's open the brackets and rewrite operator $B$ as a polynomial in $D_x$
\begin{equation}
\label{operatorB}
B=D_x^N+\alpha_1D_x^{N-1}+\dots +\alpha_N.
\end{equation}
We apply the operator $D_y$ to both sides of the first equation in \eqref{relation2} and by virtue of the conditions $\forall j\geq0 \quad\displaystyle{\frac{\partial^{j+1}}{{ \partial x^j}{\partial y}}\psi_1=0}$ we obtain the following relation
\begin{equation}
\label{relation3}
D_y(\alpha_1)D_x^{N-1}(\psi_1)+D_y(\alpha_2)D_x^{N-2}(\psi_1)+\dots +D_y(\alpha_N)\psi_1=0.
\end{equation}
Note that the functions $(\psi_1,\psi_2,\dots\psi_N)$ correspond to an arbitrary solution of the system of equations \eqref{LaxxrToda} with fixed coefficients and that's why they can be considered as independent variables. On the other hand side, a set of the variables
$(\psi_1$, $D_x^{N-1}(\psi_1)$,\dots, $D_x^{N-1}(\psi_1))$ is related with the above mentioned set of variables $(\psi_1$, $\psi_2$,\dots, $\psi_N) $ by some non-degenerate linear transformation following from the system \eqref{LaxxrToda}. For example, $\psi_{1,x}=u_{1,x}\psi_1+\psi_2$, $\psi_{1xx}=(u_{1xx}+u_{1x}^2)\psi_1+(u_{1x} +u_{2x})\psi_2+\psi_3$, etc. Thus it is also a linearly independent set. Consequently, from the equality \eqref{relation3} we have $D_y(\alpha_j)=0$ for any $j=1,2,\dots N$. It shows that the functions $\alpha_j$ are $y$-integrals of the nonlinear system \eqref{rToda}. In \cite{Smirnov15} it was proven that the functions $\alpha_1, \alpha_2,\dots \alpha_N$ form a complete set of independent $y$-integrals of the system \eqref{rToda}. Due to the invariance of the system under the involution $x\leftrightarrow y$, from this set of integrals one can easily obtain a complete set of independent $x$-integrals of the system \eqref{rToda}, as well (see also \cite{Demskoi10}).

\section{Reductions of the Ferapontov-Shabat-Yamilov lattice}

Let us concentrate on the lattice ($E6$), found by Shabat and Yamilov \cite{ShabatYamilov} and independently by Ferapontov \cite{Ferapontov97}. We consider the reductions obtained by imposing the following termination conditions
\begin{equation}
\label{FSYcutoff}
u_0=c_0, \qquad u_{m+1}=c_{m+1},
\end{equation} 
where $c_0$ and $c_{m+1}$ are arbitrary constants (the case when $u_0=0$, $u_{m+1}=\infty$ is studied earlier in \cite{Demskoi10} by an alternative method). They have the form of a system of nonlinear hyperbolic equations:
\begin{eqnarray}\label{4.1}
&&u_{1,xy}=u_{1,x}u_{1,y}\left(\frac{1}{u_1-c_0}-\frac{1}{u_{2}-u_{1}}\right),\nonumber\\
&&u_{2,xy}=u_{2,x}u_{2,y}\left(\frac{1}{u_{2}-u_{1}}-\frac{1}{u_{3}-u_{2}}\right), \nonumber \\
&&......................................... \label{rFSY}\\
&&u_{m,xy}=u_{m,x}u_{m,y}\left(\frac{1}{u_{m}-u_{m-1}}-\frac{1}{c_{m+1}-u_{m}}\right). \nonumber 
\end{eqnarray} 
It was proved in \cite{HabibullinKuznetsova20} that for an arbitrary $m\geq1$ the characteristic Lie algebras corresponding to the directions $x$ and $y$ have finite dimension that guarantees the existence of complete sets of characteristic integrals in both directions. However the integrals themselves were not found. Below we present the generating function for these integrals in terms of the Lax pair.

To begin with, we note that the first-order integral can be easily derived from elementary considerations. Indeed, let us rewrite the system in the form:
\begin{eqnarray}
&&\frac{u_{1,xy}}{u_{1,x}}=u_{1,y}\left(\frac{1}{u_1-c_0}-\frac{1}{u_{2}-u_{1}}\right),\nonumber\\
&&\frac{u_{2,xy}}{u_{2,x}}=u_{2,y}\left(\frac{1}{u_{2}-u_{1}}-\frac{1}{u_{3}-u_{2}}\right), \nonumber \\
&&......................................... \label{rFSYtransformed}\\
&&\frac{u_{m,xy}}{u_{m,x}}=u_{m,y}\left(\frac{1}{u_{m}-u_{m-1}}-\frac{1}{c_{m+1}-u_{m}}\right) \nonumber 
\end{eqnarray} 
and then adding all the equations we arrive at the relation
\begin{equation*}
D_y\log (u_{1x}u_{2x}\dots u_{mx})=D_y\log (g_1g_2\dots g_{m+1}).
\end{equation*}
It obviously follows from this fact that the function
\begin{equation}\label{J}
J=u_{1x}u_{2x}\dots u_{mx}(g_1g_2\dots g_{m+1})
\end{equation}
is a $y$-integral. Here $g_1=\frac{1}{u_1-c_0}$, $g_2=\frac{1}{u_{2}-u_{1}}$, ... $g_{m}=\frac{1}{u_{m}-u_{m-1}}$, 
$g_{m+1}=\frac{1}{c_{m+1}-u_{m}}.$
To search for higher order integrals, we will use the Lax pair, which was found by E.V. Ferapontov$\footnote{Private communication, 2020.}$:
\begin{equation}\label{FLax}
\psi_{n,x}= a_{n}(\psi_{n+1}-\psi_{n}), \quad \psi_{n,y}=b_n(\psi_{n}-\psi_{n-1}),
\end{equation} 
where $a_n=\frac{u_{n,x}}{u_{n+1}-u_{n}}$, $b_n=\frac{u_{n,y}}{u_{n}-u_{n-1}}$. By imposing on the system \eqref{FLax}, in addition to the equalities \eqref{FSYcutoff}, also conditions of the form $\psi_{0}=0$, $\psi_{m+1}=0$ we can obtain a Lax pair for system \eqref{rFSY} which is given by the following two systems of linear equations, the first one is:
\begin{eqnarray}\label{4.4}
&& \psi_{1,x}=a_{1}(\psi_2-\psi_1),\nonumber\\
&& \psi_{2,x}=a_{2}(\psi_3-\psi_2), \nonumber \\
&&......................................... \label{LaxxrFSY}\\
&&\psi_{m-1,x}=a_{m-1}(\psi_{m}-\psi_{m-1}), \nonumber \\
&&\psi_{m,x}=-a_{m}\psi_{m}. \nonumber
\end{eqnarray} 
Likewise, the second system
\begin{eqnarray}
&& \psi_{1,y}=b_1\psi_1,\nonumber\\
&& \psi_{2,y}=b_2(\psi_2-\psi_1), \nonumber \\
&&......................................... \label{LaxyrFSY}\\
&&\psi_{m-1,y}=b_{m-1}(\psi_{m-1}-\psi_{m-2}), \nonumber \\
&&\psi_{m,y}=b_m(\psi_{m}-\psi_{m-1}). \nonumber
\end{eqnarray} 
From the system \eqref{4.4} one can easily obtain a differential relation connecting the functions $\psi_1$ and $\psi_m$:
\begin{equation}\label{4.6}
\psi_m=\left(\frac{1}{a_{m-1}}D_x+1\right)\left(\frac{1}{a_{m-2}}D_x+1\right)...\left(\frac{1}{a_{1}}D_x+1\right)\psi_1.
\end{equation}
Now it is easy to see that from the last equation of the system \eqref{4.4} and  equation \eqref{4.6} the relation follows
\begin{equation}\label{4.7}
R\psi_1=0,
\end{equation}
where $R=\left(D_x+a_m\right)\left(\frac{1}{a_{m-1}}D_x+1\right)\left(\frac{1}{a_{m-2}}D_x+1\right)...\left(\frac{1}{a_{1}}D_x+1\right)\psi_1$. Therefore, the eigenfunction $\psi_1$ is a joint solution to the overdetermined system consisting of \eqref{4.7} and the equation
\begin{equation}\label{4.8}
\left(D_y-b_1\right)\psi_1=0.
\end{equation}
By virtue of the cutoff condition \eqref{FSYcutoff} the system \eqref{4.7}-\eqref{4.8} can be essentially simplified. Let us use an explicit expression for the potential $b_1$ and integrate the equation \eqref{4.8}, rewritten as follows
\begin{equation*}
\frac{d\psi_1}{\psi_1}=\frac{u_{1,y}dy}{u_1-C_0}.
\end{equation*}

As a result, we obtain an expression for the eigenfunction $\psi_1=\left(u_1-C_0\right)\varphi_1(x)$. Afterwards in \eqref{4.7}-\eqref{4.8} we move on to the new sought function $\varphi_1(x)$ and arrive at a system of the form
\begin{equation}\label{s}
\hat{R}\varphi_1=0,  \quad \frac{\partial}{\partial{y}}\varphi_1=0,
\end{equation}
where 
\begin{equation*}
\hat{R}=\left(D_x+a_m\right)\left(\frac{1}{a_{m-1}}D_x+1\right)...\left(\frac{1}{a_1}D_x+1\right)\left(u_1-C_0\right).
\end{equation*}
Repeatedly applying the ratio
\begin{equation}\label{permut}
\left(D_x+f\right)\frac{1}{g}=\frac{1}{g}\left(D_x+f-\left(\ln{g}\right)_x\right),
\end{equation}
we rewrite the operator $\hat{R}$ as follows
\begin{equation*}
\hat{R}=\frac{u_1-C_o}{a_1a_2...a_m}B,
\end{equation*}
where the operator $B$ has the form
\begin{equation}\label{operb}
B=\left(D_x+r_m\right)\left(D_x+r_{m-1}\right)...\left(D_x+r_1\right).
\end{equation}
Here $r_1=a_1+\frac{u_{1,x}}{u_1-C_0},\quad r_2=a_2-\left(\ln{a_1}\right)_x+\frac{u_{1,x}}{u_1-C_0}$, but in the general case
\begin{equation}\label{r_k} 
 r_k=a_k-\sum^{k-1}_{i=1}\left(\ln{a_i}\right)_x+\frac{u_{1,x}}{u_1-C_0}, \quad k\leq{m}.
\end{equation}
After all transformations, the system \eqref{s} converts into
\begin{equation}\label{4.10}
B\varphi_1=0, \quad \frac{\partial}{\partial{y}}\varphi_1=0.
\end{equation}
Let us represent the differential operator $B$ in the following expanded form
\begin{equation}\label{4.11}
B=D^m_x+{\beta_1}D^{m-1}_x+...+{\beta}_m.
\end{equation}
It is easy to verify, by analogy with the previous section, that the coefficients $\beta_j$ of this expansion are characteristic integrals of the system of nonlinear equations \eqref{4.1}.

The following example is really instructive.

{\bf Example 1.} Let us assume that $m=2$ in \eqref{4.1},
then we arrive at the system:
\begin{equation}\label{examplem=2}
\begin{aligned}
&\frac{u_{1,xy}}{u_{1,x}}=u_{1,y}\left(\frac{1}{u_1-C_0}-\frac{1}{u_2-u_1}\right),\\
&\frac{u_{2,xy}}{u_{2,x}}=u_{2,y}\left(\frac{1}{u_2-u_1}-\frac{1}{C_3-u_2}\right).
\end{aligned}
\end{equation}
Calculations show that the operator $B$ in this case is reduced to the following form:
\begin{equation}\label{4.12}
B=D^2_x+\left(-\frac{u_{1,xx}}{u_{1,x}}+2g_1u_{1,x}+(g_2+g_3)u_{2,x}\right)D_x+(C_3-C_0)u_{1,x}u_{2,x}g_1g_2g_3,
\end{equation}
where
\begin{equation}\label{4/12} 
g_1=\frac{1}{u_1-C_0},\quad g_2=\frac{1}{u_2-u_2}, \quad g_3=\frac{1}{C_3-u_2}.
\end{equation}
Therefore, the required integrals are
\begin{align*}
\beta_1=-\frac{u_{1,xx}}{u_{1,x}}+2g_1u_{1,x}+(g_2+g_3)u_{2,x}, \quad \beta_2=(C_3-C_0)u_{1,x}u_{2,x}g_1g_2g_3.
\end{align*}
It is worth noting that for $C_3=C_0$ the integral $\beta_2$ becomes trivial, i.e. $\beta_2=0$. In other words formula \eqref{4.11} does not always define a complete set of independent integrals, since the first-order integral $J$ given by the formula \eqref{J} might be lost.  However, even in this case, the system of equations \eqref{examplem=2} admits a complete set of $y$-integrals
\begin{equation*}
I_1=-\frac{u_{1,xx}}{u_{1,x}}+2g_1u_{1,x}+(g_2+g_3)u_{2,x}, \quad I_2=u_{1,x}u_{2,x}g_1g_2g_3.
\end{equation*}
Let us examine the structure of the set of integrals obtained from the expansion \eqref{4.11} in order to construct a complete set of integrals of the system \eqref{4.1}. As it can be seen from the formula \eqref{detdifInt}, expressing the criterion for the independence of integrals, it is necessary to clarify the form of the dependence of the integrals on the highest derivatives.
From explicit formulas \eqref{r_k} we can derive some simplified representations for the functions $r_2$, $r_3$,...,$r_m$, focusing on the dependence of these functions on the highest derivatives
\begin{equation*}
r_k=-\sum^{k-1}_{i=1}\frac{u_{i,xx}}{u_{i,x}}+h\left[u_x\right], 2\leq{k}\leq{m}.
\end{equation*} 

Here the symbol $h\left[u_x\right]$ indicates that the remaining terms can only depend on the dynamical variables $u_1, u_2, ..., u_m$ and their first-order derivatives. Our goal is to obtain similar representations for integrals $\beta_{j}$ defined in terms of the operator $B$ (see \eqref{4.11}).
Let us introduce the notation
\begin{equation}\label{Hk}
H_k:=\left(D_x+r_k\right)\left(D_x+r_{k-1}\right)...\left(D_x+r_1\right).
\end{equation}
For the operator $H_2$ the representation obviously holds 
\begin{equation*}
H_2=D^2_x-\left(\frac{u_{1,xx}}{u_{1,x}}+h\left[u_x\right]\right)D_x+h\left[u_{xx}\right].
\end{equation*}
Now we find a similar representation for $H_3$
\begin{equation*}
H_3=D^3_x-\left(2\frac{u_{1,xx}}{u_{1,x}}+\frac{u_{2,xx}}{u_{2,x}}+h\left[u_x\right]\right)D^2_x-\left(\frac{u_{1,xxx}}{u_{1,x}}+h\left[u_{xx}\right]\right)D_x+h\left[u_{xxx}\right],
\end{equation*}
where the symbol $h\left[u_{xx}\right]$ (as well as the symbol $h\left[u_{xxx}\right]$) means that the remaining terms can contain the variables $u_1, u_2,..., u_m$ and their derivatives with respect to $x$  not higher than the second order (respectively, not higher than the third order). Further, by induction we can show that for $k\leq{m}$ the representation holds
\begin{align}
H_k=D^k_x-\left((k-1)\frac{u_{1,xx}}{u_{1,x}}+(k-2)\frac{u_{2,xx}}{u_{2,x}}+...+\frac{u_{k-1,xx}}{u_{k-1,x}}+h\left[u_x\right]\right)D^{k-1}_x- \nonumber\\
\left((k-2)\frac{u_{1,xxx}}{u_{1,x}}+(k-3)\frac{u_{2,xxx}}{u_{2,x}}+...+\frac{u_{k-2,xxx}}{u_{k-2,x}}+h\left[u_{xx}\right]\right)D^{k-2}_x-\dots -  \label{4.17'}\\
\left(\frac{u_{1,[k]}}{u_{1,x}}+h\left[u_{[k-1]}\right]\right)D_x+h\left[u_{[k]}\right],\ \nonumber
\end{align}
where as before $u_{i,j}=\frac{\partial^j}{\partial{x^j}}u_i$, and the symbol $h\left[u_{[s]}\right] $ indicates that these terms can depend on the variables $u_1,u_2,...,u_m$ and on their derivatives with respect to $x$ of order no higher than $s$. Assuming $k=m$, we find a representation of the type under consideration for the operator $B=H_m$, which specifies the generating function of the characteristic integrals, from which it follows that the integrals $I_1=-\beta_{m-1}, I_2=-\beta_{m -2},...,I_{m-1}=-\beta_{1}$ admit the following representation:
\begin{eqnarray} \label{4.13}
&&I_1=\frac{u_{1,[m]}}{u_{1,x}}+h\left[u_{[m-1]}\right],\nonumber\\
&&I_2=2\frac{u_{1,[m-1]}}{u_{1,x}}+\frac{u_{2,[m-1]}}{u_{2,x}}+h\left[u_{[m-2]}\right],\nonumber\\
&&......................................... \label{4/13}\\
&&I_{m-1}=(m-1)\frac{u_{1,xx}}{u_{1,x}}+(m-2)\frac{u_{2,xx}}{u_{2,x}}+...+\frac{u_{m-1,xx}}{u_{m-1,x}}+h\left[u_x\right].\nonumber
\end{eqnarray}
Let's put $I_m=J$, where $J$ is the first-order integral given by the formula \eqref{J}. Now we summarize the arguments above as a statement.

{\bf Theorem 2}.{\it The functions $I_1, I_2,...I_m$ constitute a complete set of independent $y$-integrals of the system of equations \eqref{4.1} }

From formulas \eqref{4/13}, \eqref{J} it is easy to see that the order of the integral $I_j$ is equal to $m-j+1$. And the determinant of the form \eqref{detdifInt} corresponding to this set of integrals coincides with the determinant of the  lower-triangular matrix, having the diagonal of  the form
\begin{equation*}
\left(\frac{1}{u_{1,x}}, \frac{1}{u_{2,x}},...,\frac{1}{u_{m-1,x}},u_{1,x}u_{2,x}...u_{m-1,x}g_1g_2...g_{m+1}\right).
\end{equation*}
Thus the determinant of the matrix \eqref{detdifInt} in this case has the form $g_1g_2...g_{m+1}$ and, therefore, is different from identical zero, that proves the theorem.

Note that the system of equations \eqref{4.1} transforms into itself with a simultaneous change of variables of the form $x\longleftrightarrow{y}$, $n\longleftrightarrow{m-n+1}$, $C_0\longleftrightarrow{C_{m+1 }}$
. Therefore, from a complete set of $y$-integrals one can easily obtain a complete set of $x$-integrals.

{\bf Corollary.} {\it The system of equations \eqref{4.1} admits the complete sets of independent integrals in both characteristic directions $x$ and $y$, that is, it is integrable in the Darboux sense for any choice of the parameters $m, C_0, C_{m+1}$.}

\section{On reductions of the lattice $(E7)$}

Lattice  $(E7)$ was found in \cite{HabibullinPoptsova18} as a result of the integrable classification of nonlinear lattices. In \cite{Kuznetsova21} a Lax pair was found for this equation
\begin{equation}\label{Lax7}
\begin{aligned}
&{\psi}_{n,x}=\frac{v_{n,x}+v^2_n-1}{v_{n+1}-v_n}\left(\psi_{n+1}-\psi_n\right)-v_n\psi_n,\\
&{\psi}_{n,y}=\frac{v_{n,y}+v^2_n-1}{v_n-v_{n-1}}\left(\psi_n-\psi_{n-1}\right)-v_n\psi_n.
\end{aligned}
\end{equation}
In other words system \eqref{Lax7} is consistent if and only if the function $v_n=v_{n}(x,y)$ solves equation $(E7)$.

Let's consider a finite-field system of equations of hyperbolic type 
\begin{equation}\label{red7}
\begin{aligned}
&v_{1,xy}={\alpha}_1\left(v_{1,x}+v^2_1-1\right)\left(v_{1,y}+v^2_1-1\right)-2v_1\left(v_{1,x}+v_{1,y}+v^2_1-1\right),\\
&v_{k,xy}=\alpha_k\left(v_{k,x}+v^2_k-1\right)\left(v_{k,y}+v^2_k-1\right)-2v_k\left(v_{k,x}+v_{k,y}+v^2_k-1\right), \\
&v_{m,xy}={\alpha}_m\left(v_{m,x}+v^2_m-1\right)\left(v_{m,y}+v^2_m-1\right)-2v_m\left(v_{m,x}+v_{m,y}+v^2_m-1\right)
\end{aligned}
\end{equation}
obtained from the lattice $(E7)$ by imposing the following termination conditions:
\begin{equation}\label{cutoff7}
v_0=\epsilon,\qquad v_{m+1}=\delta.
\end{equation}
Here $ 2\leq{k}\leq{m-1}$ and
\begin{equation}\nonumber
\begin{aligned}
&\alpha_1=\frac{1}{v_1-\epsilon}-\frac{1}{v_2-v_1},\\
&\alpha_k=\frac{1}{v_k-v_{k-1}}-\frac{1}{v_{k+1}-v_k},\\
&\alpha_m=\frac{1}{v_m-v_{m-1}}-\frac{1}{\delta-v_m}.
\end{aligned}
\end{equation}
Note that each of the parameters $\epsilon$ and $\delta$ can take only one of two values $\epsilon=\pm1,\, \delta=\pm1$. We emphasize that in this case the linear system of equations \eqref{Lax7} allows us to obtain the Lax pair for the reduced system \eqref{red7}. To do this, in addition to the conditions \eqref{cutoff7} for cutting off the chain, it is necessary to impose conditions for terminating the Lax pair:
\begin{equation}\label{cutoffLax7}
\psi_0=0,\qquad \psi_{m+1}=0.
\end{equation}
As a result, we arrive at the linear problem in $x$:
\begin{eqnarray}
&&\psi_{1,x}=\frac{v_{1,x}+v^2_1-1}{v_2-v_1}\left(\psi_2-\psi_1\right)-v_1\psi_1\nonumber\\
&&..........................................\label{5.5}\\
&&\psi_{m-1,x}=\frac{v_{m-1,x}+v^2_{m-1}-1}{v_m-v_{m-1}}\left(\psi_m-\psi_{m-1}\right)-v_{m-1}\psi_{m-1},\nonumber\\
&&\psi_{m,x}=\left(\frac{v_{m,x}}{v_m-\delta}+\delta\right)\psi_m,\nonumber
\end{eqnarray}
 
and the linear problem in $y$

\begin{eqnarray}
&&\psi_{1,y}=\left(\frac{v_{1,y}}{v_1-\epsilon}+\epsilon\right)\psi_1,\nonumber\\
&&\psi_{2,y}=\frac{v_{2,y}+v^2_2-1}{v_2-v_1}\left(\psi_2-\psi_1\right)-v_2\psi_1\nonumber\\
&&.....................................................\label{5.6}\\
&&\psi_{m,y}=\frac{v_{m,y}+v^2_m-1}{v_m-v_{m-1}}\left(\psi_m-\psi_{m-1}\right)-v_m\psi_{m-1}.\nonumber
\end{eqnarray}
For brevity, we introduce the following notation
\begin{equation*}
A_j=\frac{v_{j,x}+v^2_j-1}{v_{j+1}-v_j}
\end{equation*}
and using the system of equations \eqref{5.5}, we express the function $\psi_m$ in terms of $\psi_1$:
\begin{equation}\label{5.7}
\psi_m=\frac{1}{A_{m-1}}\left(D_x+B_{m-1}\right)\frac{1}{A_{m-2}}\left(D_x+B_{m-2}\right)...\frac{1}{A_1}\left(D_x-B_1\right)\psi_1,
\end{equation}
where $B_j=-A_j-v_j$. Let us rewrite the last equation of the system \eqref{5.5} in the form
\begin{equation}\label{5.8}
\left(D_x-B_m\right)\psi_m=0,\quad \mbox{where}\quad B_m=\frac{v_{m,x}}{v_m-\delta}+\delta.
\end{equation}
From equalities \eqref{5.7}, \eqref{5.8} it obviously follows that the function $\psi_1$ is a solution to the linear differential equation of order $m$:
\begin{equation}\label{5.9}
\left(D_x-B_m\right)\frac{1}{A_{m-1}}\left(D_x-B_{m-1}\right)...\frac{1}{A_1}\left(D_x-B_1\right)\psi_1=0.
\end{equation}
Moreover, from the system of equations \eqref{5.6} we have one more equation for $\psi_1$:
\begin{equation}\label{5.10}
\left(D_y-\frac{v_{1,y}}{v_1-\epsilon}-\epsilon\right)\psi_1=0.
\end{equation}
By replacing the variables $\psi_1=\left(v_1-\epsilon\right)e^{\epsilon{y}}\varphi_1$ we reduce the system of equations \eqref{5.9}, \eqref{5.10} to a simpler form
\begin{equation}\label{5.x}
\frac{\partial}{\partial{y}}\varphi_1=0, \qquad R\varphi_1=0,
\end{equation}
where
\begin{equation}\label{5.11}
R=\left(D_x-B_m\right)\frac{1}{A_{m-1}}\left(D_x-B_{m-1}\right)...\frac{1}{A_1}\left(D_x-B_1\right)\left(v_1-\epsilon\right),
\end{equation}
since the factor $e^{\epsilon{y}}$ commutes with $R$. Using the permutation formula \eqref{permut}, we can show that the equalities are satisfied 
\begin{eqnarray}
&&\left(D_x-B_1\right)\left(v_1-\epsilon\right)=\left(v_1-1\right)\left(D_x+\left(\log(v_1-\epsilon)\right)_x-B_1\right),\nonumber\\
&&\left(D_x-B_2\right)\frac{1}{A_1}\left(D_x-B_1\right)\left(v_1-\epsilon\right)=\nonumber\\ 
&&=\frac{v_1-\epsilon}{A_1}\left(D_x+\left(\log\frac{v_1-\epsilon}{A_1}\right)_x-B_2\right)\left(D_x+\left(\log(v_1-\epsilon)\right)_x-B_1\right),\nonumber
\end{eqnarray}
etc. It is easy to prove by induction that the operator $R$ can be reduced to the form
\begin{equation}\label{5.12}
R=\frac{v_1-\epsilon}{A_1A_2...A_{m-1}}\left(D_x+r_m\right)\left(D_x+r_{m-1}\right)...\left(D_x+r_1\right)
\end{equation}
where
\begin{equation}\label{rk}
r_k=\left(\log\frac{v_1-\epsilon}{A_1A_2...A_{k-1}}\right)_x-B_k.
\end{equation}
As a result, the system of equations \eqref{5.x} above takes the form:
\begin{equation}\label{5.14}
\frac{\partial}{\partial{y}}\varphi_1=0, \qquad L\varphi_1=0,
\end{equation}
where
\begin{equation*}
L=\left(D_x+r_m\right)\left(D_x+r_{m-1}\right)...\left(D_x+r_1\right).
\end{equation*}
Let us present the operator $L$ in an expanded form
\begin{equation}\label{operL}
L=D^m_x+\beta_1D^{m-1}_x+...+\beta_m.
\end{equation}
Using arguments similar to those given above in Section 3, we can use the system \eqref{5.14} to show that the coefficients $\beta_1, \beta_2,...,\beta_m$ are $y$-integrals of the system of equations \eqref{red7}.
In order to construct a complete set of independent $y$-integrals of the system \eqref{red7} we need the following auxiliary statement.

{\bf Lemma 1}.
{\it The function 
\begin{equation}\label{Int}
J=\frac{\left(v_{1,x}+v^2_1-1\right)\left(v_{2,x}+v^2_2-1\right)...\left(v_{m,x}+v^2_m-1\right)}{\left(v_1-\epsilon\right)\left(v_2-v_1\right)\left(v_3-v_2\right)...\left(v_m-v_{m-1}\right)\left(v_m-\delta\right)}e^{-\left(\epsilon+\delta\right)y}
\end{equation}
is a  first order $y$-integral of the system of equations \eqref{red7}.}

The lemma is proved by a simple but cumbersome calculation, so we present it in the specific case $m=3$.
Consider the system \eqref{red7} for $m=3$ and reduce it to the following form:
\begin{equation}\label{red7lem}
\begin{aligned}
&D_y\log\left(v_{1,x}+v^2_1-1\right)=\frac{v_{1,y}}{v_1-\epsilon}-v_1+\epsilon-\frac{v_{1,y}+v^2_1-1}{v_2-v_1}\\
&D_y\log\left(v_{2,x}+v^2_2-1\right)=\frac{v_{2,y}+v^2_2-1}{v_2-v_1}-2v_2-\frac{v_{2,y}+v^2_2-1}{v_3-v_2}
\\
&D_y\log\left(v_{3,x}+v^2_3-1\right)=\frac{v_{3,y}+v^2_3-1}{v_3-v_2}-2v_3-\frac{v_{3,y}+v^2_3-1}{\delta-v_3}.
\end{aligned}
\end{equation}
Let's add up all the equations of the system \eqref{red7lem} and get the relation
\begin{equation*}
\begin{aligned}
&D_y\log\left(v_{1,x}+v^2_1-1\right)\left(v_{2,x}+v^2_2-1\right)\left(v_{3,x}+v^2_3-1\right)=\\
&D_y\log\left(v_1-\epsilon\right)\left(v_2-v_1\right)\left(v_3-v_2\right)\left(v_3-\delta\right)+\epsilon+\delta,
\end{aligned}
\end{equation*}
from which it follows that
\begin{equation}\label{lem}
D_y\log\frac{\left(v_{1,x}+v^2_1-1\right)\left(v_{2,x}+v^2_2-1\right)\left(v_{3,x}+v^2_3-1\right)}{\left(v_1-\epsilon\right)\left(v_2-v_1\right)\left(v_3-v_2\right)\left(v_3-\delta\right)}=\epsilon+\delta.
\end{equation}
Now it is very easy to get the statement of the lemma.

Consider the following set of $y$-integrals of the system of equations \eqref{red7}:
\begin{equation}\label{nabor}
I_1:=-\beta_{m-1}, I_2:=\beta_{m-2},...I_{m-1}=-\beta_1, I_m=J,
\end{equation}
where the functions $\beta_j$ coincide with the expansion coefficients \eqref{operL} of the operator $L$, and the integral $J$ is given by \eqref{Int}.

{\bf Theorem 3.}
{\it The functions $I_1, I_2,...,I_m$ form a complete set of independent $y$-integrals of the system of equations \eqref{red7}.}

To prove the theorem, it is enough to verify that the determinant of the matrix constructed according to the rule
\begin{equation}\label{matr}
\begin{vmatrix}
\frac{\partial I_1}{\partial u_{1,[m]}} & \frac{\partial I_1}{\partial u_{2,[m]}} & \ldots & \frac{\partial I_1}{\partial u_{m,[m]}}\\
\frac{\partial I_2}{\partial u_{1,[m-1]}} & \frac{\partial I_2}{\partial u_{2,[m-1]}} & \ldots & \frac{\partial I_2}{\partial u_{m,[m-1]}}\\
\ldots&\ldots&\ldots&\ldots\\
  \frac{\partial I_m}{\partial u_{1,x}} & \frac{\partial I_m}{\partial u_{2,x}} & \ldots & \frac{\partial I_m}{\partial u_{m,x}}
\end{vmatrix}
\end{equation}
differs from identical zero. To do this, it is necessary to find out the dependence of the integrals on the higher order derivatives of the dynamical variables included in them. From explicit representations \eqref{rk} for functions $r_k$ we have
\begin{equation}\label{rkk}
r_k=-\frac{v_{1,xx}}{v_{1,x}+v^2_1-1}-\frac{v_{2,xx}}{v_{2,x}+v^2_2-1}-...-\frac{v_{k-1,xx}}{v_{k-1,x}+v^2_{k-1}-1}+h[v_x],
\end{equation}
where $2\leq{k}\leq{m}$. Here the symbol $h[v_x]$ means that remaining terms contain dynamical variables and their first order derivatives. By analogy with the formula \eqref{4.17'} the operator $L$ can be represented as:
\begin{equation*}
L=D^m_x-\left(\frac{(m-1)v_{1,xx}}{v_{1,x}+v^2_1-1}+\frac{(m-2)v_{2,xx}}{v_{2,x}+v^2_2-1}-...-\frac{v_{k-1,xx}}{v_{k-1,x}+v^2_{k-1}-1}+h[v_x]\right)D^{m-1}_x-
\end{equation*}
\begin{equation*}
-\left(\frac{(m-2)v_{1,xxx}}{v_{1,x}+v^2_1-1}+\frac{(m-3)v_{2,xxx}}{v_{2,x}+v^2_2-1}+...+\frac{v_{m-2,xxx}}{v_{m-2,x}}+h[v_{xx}]\right)D^{k-2}_x-...-
\end{equation*}
\begin{equation*}
-\left(\frac{v_{1,[m]}}{v_{1,x}+v^2_1-1}+h[v_{[k-1]}]\right)D_x+h[v_{[k]}],
\end{equation*}
from which it follows that the integrals can be represented in the following form, expressing the dependence of the integrals on the highest order derivatives
\begin{equation*}
\begin{aligned}
&I_1=\frac{v_{1,[m]}}{v_{1,x}+v^2_1-1}+...\\
&I_2=\frac{2v_{1,[m-1]}}{v_{1,x}+v^2_1-1}+\frac{v_{2,[m-1]}}{v_{2,x}+v^2_2-1}+...\\
&..................\\
&I_{m-1}=\frac{(m-1)v_{1,xx}}{v_{1,x}+v^2_1-1}+\frac{(m-2)v_{2,xx}}{v_{2,x}+v^2_2-1}+...+\frac{v_{m-1,xx}}{v_{m-1,x}}+... .
\end{aligned}
\end{equation*}
Recall that $I_m=J$ is given explicitly by the equality \eqref{Int}. Now to complete the proof of the theorem we have to note that the determinant of the matrix \eqref{matr} is equal to the expression:
\begin{equation*}
\frac{e^{-(\epsilon+\delta)y}}{\left(v_1-\epsilon\right)\left(v_2-v_1\right)\left(v_3-v_2\right)...\left(v_m-v_{m-1}\right)\left(v_m-\delta\right)}.
\end{equation*}
Below we present an illustrative example assuming $m=2$.

{\bf Example 2.} For $m=2$ the system \eqref{red7} takes the form
\begin{equation}\label{sys2}
\begin{aligned}
&v_{1,xy}=\left(\frac{1}{v_1-\epsilon}-\frac{1}{v_2-v_1}\right)\left(v_{1,x}+v^2_1-1\right)\left(v_{1,y}+v^2_1-1\right)-2v_1\left(v_{1,x}+v_{1,y}+v^2_1-1\right)\\
&v_{2,xy}=\left(\frac{1}{v_2-v_1}-\frac{1}{\delta-v_2}\right)\left(v_{2,x}+v^2_2-1\right)\left(v_{2,y}+v^2_2-1\right)-2v_2\left(v_{2,x}+v_{2,y}+v^2_2-1\right),
\end{aligned}
\end{equation}

The corresponding operator $L$ is represented as follows
\begin{equation}\label{prm2}
L=\left(D_x+r_2\right)\left(D_x+r_1\right)=D^2_x+\left(r_1+r_2\right)D_x+r_1r_2+r_{1,x},
\end{equation}
where
\begin{equation}\label{r1}
r_1=\frac{v_{1,x}}{v_1-\epsilon}-B_1=\frac{v_{1,x}}{v_1-\epsilon}+\frac{v_{1,x}}{v_2-v_1}-\frac{1-v_1v_2}{v_2-v_1} 
\end{equation}
and
\begin{equation}\label{r2}
r_2=\left(\log\frac{v_1-\epsilon}{A_1}\right)_x-B_2=\frac{v_{1,x}}{v_1-\epsilon}-\frac{v_{1,xx}+2v_1v_{1,x}}{v_{1,x}+v^2_1-1}+\frac{v_{2,x}+v_2\delta-1}{\delta-v_2}+\frac{v_{2,x}-v_{1,x}}{v_2-v_1}.
\end{equation}                                                                                                                   
The integral $\beta_1=I$ is easily evaluated:
\begin{equation*}
I=r_1+r_2=\frac{2v_{1,x}}{v_1-\epsilon}-\frac{v_{1,xx}+2v_1v_{1,x}}{v_{1,x}+v^2_1-1}+\frac{v_{2,x}+v_1v_2-1}{v_2-v_1}+\frac{v_{2,x}+v_2\delta-1}{\delta-v_2}.
\end{equation*}
Due to the Lemma 1 we have the integral $\beta_2=J$ (evidently it is non-autonomous for $\epsilon\neq-\delta$):
\begin{equation}
J=\frac{\left(v_{1,x}+v^2_1-1\right)\left(v_{2,x}+v^2_2-1\right)}{\left(v_1-\epsilon\right)\left(v_2-v_1\right)\left(v_2-\delta\right)}e^{-\left(\epsilon+\delta\right)y}.
\end{equation}
These two functions $I$ and $J$ constitute the complete set of the $y$-integrals for the system \eqref{sys2}.

\section{On the reductions of the two-dimensional Volterra equation}

It is well known that the two-dimensional Volterra equation
\begin{equation}\label{Volt}
\begin{aligned}
&a_{n,y}=a_n\left(b_n-b_{n+1}\right),\\
&b_{n,x}=b_n\left(a_{n-1}-a_n\right)
\end{aligned}
\end{equation}
admits a Lax pair of the following form (see, for instance, \cite{ShabatYamilov},\cite{Ferapontov97}):
\begin{equation}\label{laxVolt}
\psi_{n,x}=a_n\left(\psi_{n+1}-\psi_n\right), \qquad\psi_{n,y}=b_n\left(\psi_n-\psi_{n-1}\right).
\end{equation}
By imposing the boundary conditions $a_0=0,\, b_{m+1}=0$ on the lattice \eqref{Volt}, we arrive at a finite system of hyperbolic equations
\begin{equation}\label{Volt1}
\begin{aligned}
&a_{1,y}=a_1\left(b_1-b_2\right), \\
&a_{2,y}=a_2\left(b_2-b_3\right), \\
&.................................\\
&a_{m,y}=a_mb_m, \\
&b_{1,x}=-b_1a_1,\\
&b_{2,x}=b_2\left(a_1-a_2\right),\\
&.................................\\
&b_{m,x}=b_m\left(a_{m-1}-a_m\right),
\end{aligned}
\end{equation} 
which is also integrable (see\cite{LSS81}). Its Lax pair can be easily derived from the linear system \eqref{laxVolt} by setting $\psi_0=0$, $\psi_{m+1}=0$. Then the linear system in $x$ takes the form:
\begin{equation}\label{linx}
\begin{aligned}
&\psi_{1,x}=a_1\left(\psi_2-\psi_1\right),\\
&\psi_{2,x}=a_2\left(\psi_3-\psi_2\right),\\
&...............\\
&\psi_{m,x}=-a_m\psi_m.
\end{aligned}
\end{equation}
In a similar way we get the linear system in $y$:
\begin{equation}\label{liny}
\begin{aligned}
&\psi_{1,y}=b_1\psi_1,\\
&\psi_{2,y}=b_2\left(\psi_2-\psi_1\right),\\
&......................\\
&\psi_{m,y}=b_m\left(\psi_m-\psi_{m-1}\right).
\end{aligned}
\end{equation}
It is worth noting that the system \eqref{linx}, \eqref{liny} exactly goes into the system \eqref{LaxxrFSY}-\eqref{LaxyrFSY}, studied in section 4, if the relations are satisfied
\begin{equation}\label{sootn}
a_n=\frac{u_{n,x}}{u_{n+1}-u_n},\quad b_n=\frac{u_{n,y}}{u_n-u_{n-1}},
\end{equation}
as well as boundary conditions \eqref{FSYcutoff}. The matter is that the relations \eqref{sootn} implement the Miura-type transformation between equations $(E6)$ and \eqref{Volt}. Let's rewrite some of the equations \eqref{Volt1} in the following form:
\begin{equation}\label{ur}
\begin{aligned}
&\frac{a_{1,y}}{a_1}=b_1-b_2,\\
&\frac{a_{2,y}}{a_2}=b_2-b_3,\\
&..........\\
&\frac{a_{m,y}}{a_m}=b_m.
\end{aligned}
\end{equation}
Then, summing these equalities term by term, we obtain $D_y \log{a_1a_2...a_m}=b_1$. Further, keeping in mind that due to \eqref{FSYcutoff} and \eqref{sootn} the equality $b_1=D_y \log\left(u_1-C_0\right)$ holds we find  $D_y \log\left(u_1-C_0\right)=D_y \log{a_1a_2...a_m}$. Afterwards by integrating, we find
\begin{equation}\label{u1}
u_1=C_0+a_1a_2...a_mp(x),
\end{equation}
where $p(x)$ is an arbitrary function.

Now, to construct the integrals of the system \eqref{Volt1}, one can use the scheme outlined in Section 4, where the key role is played by the operator $B$ \eqref{operb}. To do this, in explicit representations \eqref{r_k} for the coefficients $r_k$ it is necessary to exclude the variable $u_1$ due to \eqref{u1}. Simple calculations lead to the following expressions
 
\begin{equation}\label{rkvolt}
r_k=a_k+\sum^m_{i=k}\frac{a_{i,x}}{a_i}+\frac{p'(x)}{p(x)}, k=1,2,...m.
\end{equation}
For the coefficients of the operator \eqref{Hk} we find convenient representations in order to clarify their dependence on the higher order derivatives of the dynamical variables $a_1,a_2,...a_m$. For example, for $H_2$ we have
\begin{equation*}
H_2=\left(D_x+\sum^m_{i=2}\frac{a_{i,x}}{a_i}+h[a]\right)\left(D_x+\sum^m_{i=1}\frac{a_{i,x}}{a_i}+h[a]\right)=
\end{equation*}
\begin{equation*}
D^2_x+\left(\sum^m_{i=2}\frac{a_{i,x}}{a_i}+\sum^m_{i=1}\frac{a_{i,x}}{a_i}+h[a]\right)D_x+\sum^m_{i=1}\frac{a_{i,xx}}{a_i}+h[a_x]
\end{equation*}
\begin{equation*}
H_3=D^3_x+\left(\sum^m_{i=3}\frac{a_{i,x}}{a_i}+\sum^m_{i=2}\frac{a_{i,x}}{a_i}+\sum^m_{i=1}\frac{a_{i,x}}{a_i}+h[a]\right)D^2_x+
\end{equation*}
\begin{equation*}
\left(\sum^m_{i=2}\frac{a_{i,xx}}{a_i}+\sum^m_{i=1}\frac{a_{i,xx}}{a_i}+h[a_x]\right)D_x+\sum^m_{i=1}\frac{a_{i,xxx}}{a_i}+h[a_{xx}].
\end{equation*}
Here the symbols $h[a], h[a_x]$ and $h[a_{xx}]$ indicate that the terms located at this place may depend on $x$, the dynamical variables $a_1, a_2 , ...a_m$ themselves and their derivatives with respect to $x$ of the order no higher than 2. By induction it can be shown that the operator $H_m$, coinciding with the operator $B$ \eqref{operb}, admits the following representation:
\begin{equation}\label{Hm}
\begin{aligned}
&H_m=D^m_x+\left(\frac{a_{m,x}}{a_m}+\sum^m_{i=m-1}\frac{a_{i,x}}{a_i}+\sum^m_{i=m-2}\frac{a_{i,x}}{a_i}+...\sum^m_{i=1}\frac{a_{i,x}}{a_i}+h[a]\right)D^{m-1}_x+\\
&\left(\sum^m_{i=m-1}\frac{a_{i,xx}}{a_i}+\sum^m_{i=m-2}\frac{a_{i,xx}}{a_i}+...\sum^m_{i=1}\frac{a_{i,xx}}{a_i}+h[a_x]\right)+...+\sum^m_{i=1}\frac{a_{i,[m]}}{a_i}+h[a_{[m-1]}],
\end{aligned}
\end{equation}
where $a_{i,[s]}=\frac{\partial^s}{\partial{x^s}}a_i$, and the symbol $h[a_{[s]}]$ indicates that the terms located in this place depend on $x$, on the dynamical variables $a_1, a_2,...a_m$ and on their derivatives with respect to $x$ of order no higher than $s$.

From the latter we find some special representation modulo the lower order derivatives for the coefficients of the operator $H_m=B$:
\begin{equation*}
H_m=D^m_x+\beta_1D^{m-1}_x+\beta_2D^{m-2}_x+...+\beta_m,
\end{equation*}
which are $y$-integrals of the system of equations \eqref{Volt1},
\begin{equation}\label{intVolt}
\begin{aligned}
&\beta_1=\frac{a_{m,x}}{a_m}+\sum^m_{i=m-1}\frac{a_{i,x}}{a_i}+...+\sum^m_{i=1}\frac{a_{i,x}}{a_i}+...\\
&\beta_2=\sum^m_{i=m-1}\frac{a_{i,xx}}{a_i}+\sum^m_{i=m-2}\frac{a_{i,xx}}{a_i}+...+\sum^m_{i=1}\frac{a_{i,xx}}{a_i}+...\\
&..........................\\
&\beta_{m-1}=\sum^m_{i=2}\frac{a_{i,[m-1]}}{a_i}+\sum^m_{i=1}\frac{a_{i,[m-1]}}{a_i}+...\\
&\beta_m=\sum^m_{i=1}\frac{a_{i,[m]}}{a_i}+... .
\end{aligned}
\end{equation}
Now we  calculate the determinant of a matrix of the form \eqref{detdifInt} for the integrals $I_j=\beta_j$. Let's first consider a simple example, assuming $m=2$. For $m=2$ we have
\begin{equation}\label{m2}
\begin{aligned}
&\frac{\partial\beta_1}{\partial{a_{1,x}}}=\frac{1}{a_1},\\
&\frac{\partial\beta_1}{\partial{a_{2,x}}}=\frac{2}{a_2},\\
&\frac{\partial\beta_2}{\partial{a_{1,xx}}}=\frac{1}{a_1},\\
&\frac{\partial\beta_2}{\partial{a_{2,xx}}}=\frac{1}{a_2}.
\end{aligned}
\end{equation}
The required determinant has the form:
\begin{equation}\label{opr}
\begin{vmatrix}
\frac{1}{a_1}&\frac{2}{a_2}\\
\frac{1}{a_1}&\frac{1}{a_2}
\end{vmatrix}=-\frac{1}{a_1a_2}.
\end{equation}
Similarly, for $m=3$ one can check that the determinant equals $-\frac{1}{a_1a_2a_3}$. In the case of an arbitrary $m$, the determinant of the form \eqref{detdifInt} coincides with the determinant of the matrix
\begin{equation}\label{porm}
\begin{vmatrix}
\frac{1}{a_1}&\frac{2}{a_2}&\frac{3}{a_3}&...&\frac{m-2}{a_{m-2}}&\frac{m-1}{a_{m-1}}&\frac{m}{a_m}\\
\frac{1}{a_1}&\frac{2}{a_2}&\frac{3}{a_3}&...&\frac{m-2}{a_{m-2}}&\frac{m-1}{a_{m-1}}&\frac{m-1}{a_m}\\
.............&.............&.............&...&...................&...................&..........\\
\frac{1}{a_1}&\frac{2}{a_2}&\frac{2}{a_3}&...&\frac{2}{a_{m-2}}&\frac{2}{a_{m-1}}&\frac{2}{a_m}\\
\frac{1}{a_1}&\frac{1}{a_2}&\frac{1}{a_3}&...&\frac{1}{a_{m-2}}&\frac{1}{a_{m-1}}&\frac{1}{a_m}\\
\end{vmatrix}
\end{equation}
By simple linear transformations, namely, subtracting the second row from the first one, the third row from the second one, etc., the matrix \eqref{porm} can be reduced to a triangular form and make sure that its determinant is equal to $-\frac{1}{a_1a_2. ..a_m}$. Consequently, the set of $y$ -integrals \eqref{intVolt} forms a complete set of independent integrals.

{\bf Example 3.} For $m=2$ the system \eqref{Volt1} takes the form

\begin{equation}\label{Voltm=2}
\begin{aligned}
&a_{1,y}=a_1\left(b_1-b_2\right), \\
&a_{2,y}=a_2b_2, \\
&b_{1,x}=-b_1a_1,\\
&b_{2,x}=b_2\left(a_1-a_2\right).
\end{aligned}
\end{equation} 

Let's find the complete set of the $y$-integrals for this system. In this case the functions $r_1$ and $r_2$ are given by
\begin{eqnarray} 
&&r_1=a_1+\frac{a_{1,x}}{a_1}+\frac{a_{2,x}}{a_2}+\frac{p'(x)}{p(x)},\label{firstone} \\
&&r_2=a_2+\frac{a_{2,x}}{a_2}+\frac{p'(x)}{p(x)}.  \label{secondone}
\end{eqnarray}
Without loss of generality we can put $p(x)=1$. For the operator $H_2$ we have the following representation:
\begin{equation}
H_2=D^2_x+(r_1+r_2)D_x+r_1r_2+r_{1,x},
\end{equation}
where functions $\beta_1=r_1+r_2$ and $\beta_2=r_1r_2+r_{1,x}$ are the desired integrals. In virtue of the formulas \eqref{firstone}, \eqref{secondone}
we have the complete set of $y$-integrals for the system \eqref{Voltm=2}
\begin{eqnarray}
&\beta_1&=a_2+a_1+2\frac{a_{2,x}}{a_2}+\frac{a_{1,x}}{a_1}, \nonumber\\
&\beta_2&=\frac{a_{1,xx}}{a_1}+\frac{a_{2,xx}}{a_2}+a_1a_2+a_{2,x}+a_{1,x}+\frac{a_2a_{1,x}}{a_1}+\frac{a_1a_{2,x}}{a_2}+
+\frac{a_{2,x}a_{1,x}}{a_2a_1}-\left(\frac{a_{1,x}}{a_1}\right)^2.\nonumber
\end{eqnarray}

\end{document}